\documentclass{emulateapj}
\submitted{Accepted by ApJL}

\begin{document}
\title{High Resolution X-ray Imaging of Supernova Remnant 1987A}

\shorttitle{HRC Observations of SNR 1987A}
\shortauthors{Ng et al.}

\author{C.-Y. Ng\altaffilmark{1},
B.~M. Gaensler\altaffilmark{1},
S.~S. Murray\altaffilmark{2},
P.~O. Slane\altaffilmark{2},
S.~Park\altaffilmark{3},
L.~Staveley-Smith\altaffilmark{4},
R.~N. Manchester\altaffilmark{5},
D.~N. Burrows\altaffilmark{3}}
\altaffiltext{1}{Sydney Institute for Astronomy, School of Physics, The University of Sydney, NSW 2006, Australia}
\altaffiltext{2}{Harvard-Smithsonian Center for Astrophysics, Cambridge, MA 02138, USA}
\altaffiltext{3}{Department of Astronomy and Astrophysics, Pennsylvania State University, University Park, PA 16802, USA}
\altaffiltext{4}{International Centre for Radio Astronomy Research, School of Physics, The University of Western Australia, Crawley, WA 6009, Australia}
\altaffiltext{5}{Australia Telescope National Facility, CSIRO, Marsfield, NSW 1710, Australia}
\email{ncy@physics.usyd.edu.au}

\begin{abstract}
We report observations of the remnant of Supernova 1987A with the High
Resolution Camera (HRC) onboard the \emph{Chandra X-ray Observatory}. A direct
image from the HRC resolves the annular structure of the X-ray remnant,
confirming the morphology previously inferred by deconvolution of lower
resolution data from the Advanced CCD Imaging Spectrometer. Detailed spatial
modeling shows that the a thin ring plus a thin shell gives statistically the
best description of the overall remnant structure, and suggests an outer radius
$0.96\arcsec\pm0.05\arcsec \pm0.03\arcsec$ for
the X-ray--emitting region, with the two uncertainties corresponding to the
statistical and systematic errors, respectively. This is very similar to the
radius determined by a similar modeling technique for the radio shell at a
comparable epoch, in contrast to previous claims that the remnant is
10-15\% smaller at X-rays than in the radio band. The HRC observations
put a flux limit of 0.010\,cts\,s$^{-1}$ (99\% confidence level, 0.08-10\,keV
range) on any compact source at the remnant center. Assuming the same foreground
neutral hydrogen column density as towards the remnant, this allows us to rule
out an unobscured neutron star with surface temperature $T^\infty>2.5$\,MK
observed at infinity, a bright pulsar wind nebula or a magnetar.
\end{abstract}


\section{Introduction}
The core-collapse supernova \object{(SN) 1987A} in the Large Magellanic
Cloud was the brightest SN observed since the invention of modern
telescopes, providing the best opportunity to study the last evolutionary
stage of a massive star. Optical observations have revealed a triple-ring
nebula centered on the explosion, consisting of an inner equatorial
ring of radius 0.81\arcsec-0.86\arcsec and two larger outer rings
\citep{bkh+95,plc+95}. These are part of the hourglass-shaped circumstellar
medium (CSM) ejected by the progenitor 20,000 years ago \citep[see][]{mp07}.
Since early 2004, the SN blast wave has begun to encounter the main body of
inner ring \citep{pzb+05,pzb+06}. This `big crash' has led to a drastic
increase in soft X-ray emission that originates from the optically thin
thermal plasma behind the shock. The X-ray spectrum is well-described by a
two-component plane-parallel shock model in nonequilibrium ionization,
with plasma temperatures $kT\sim 2$ and 0.3\,keV, corresponding
to the fast and decelerated shocks, respectively \citep{pzb+04,pzb+06}. As
the blast wave propagates into the dense CSM, the soft X-ray emission traces
the evolution of the forward shock, probing the CSM structure and its density
profile.

The nearness of SN~1987A (51.4\,kpc) allows us to resolve a supernova remnant
(SNR) morphology at a very young age. In X-rays, this is only possible with
the \emph{Chandra X-ray Observatory}, the highest resolution X-ray telescope
compared to any previous, current and even planned future X-ray missions.
Previous \emph{Chandra} observations of SNR~1987A were all carried out by
the Advanced CCD Imaging Spectrometer (ACIS). However, the ACIS detector has
a pixel size 0.492\arcsec, comparable to the full width at half maximum
(FWHM) of the mirror's on-axis point-spread function (PSF). Although the
effective resolution can be slightly improved by the dithering of the
spacecraft and by applying a sub-pixel imaging algorithm \citep{tmm+01},
image deconvolution is still required to fully resolve the remnant structure
\citep[e.g.][]{bmh+00}. To determine whether artifacts are introduced by the
complicated non-linear reconstruction process, the ACIS results need to be
compared with higher resolution direct X-ray images. An analogy can be drawn
from the radio imaging campaign of SNR~1987A. Since 1992, Australia Telescope
Compact Array (ATCA) observations at 9\,GHz have revealed detailed structure
of the radio shell using the super-resolved technique, but it was not until
the upgrade of the ATCA in 2003 that the higher resolution diffraction-limited
images at 18\,GHz provided a direct image of the radio morphology
\citep[see review by][]{gsm+07}.

The High Resolution Camera (HRC) onboard \emph{Chandra}, consisting of two
microchannel plate detectors, offers smaller electronic readout pixels
(0.13175\arcsec) than ACIS. These better sample the PSF, providing a
more straightforward imaging process without the need for deconvolution.
Despite a smaller effective area for HRC than ACIS and the lack of any
spectral resolution, the SNR is now very bright in X-rays below 2\,keV,
at which the HRC has good sensitivity, making it an ideal instrument for
morphological studies. In this Letter, we report a detailed analysis of the
first HRC observation of SNR~1987A.

\begin{figure*}[ht]
\centering
 \begin{minipage}[c]{0.3\textwidth}
  \includegraphics[height=53mm, bb=5 49 544 503, clip=]{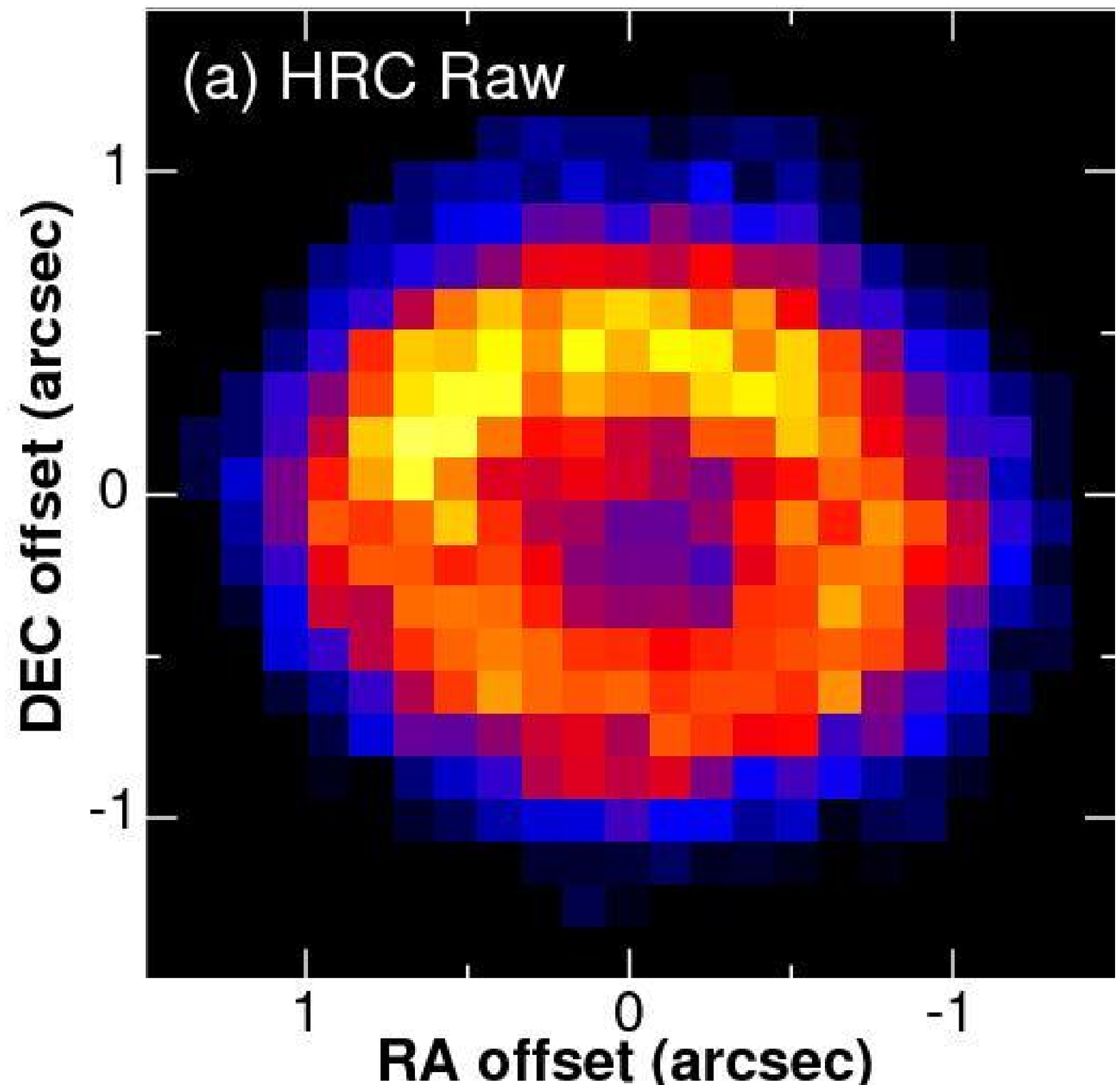}
 \end{minipage}
\hspace*{6.0mm}
 \begin{minipage}[c]{0.3\textwidth}
  \includegraphics[height=53mm, bb=89 49 544 503, clip=]{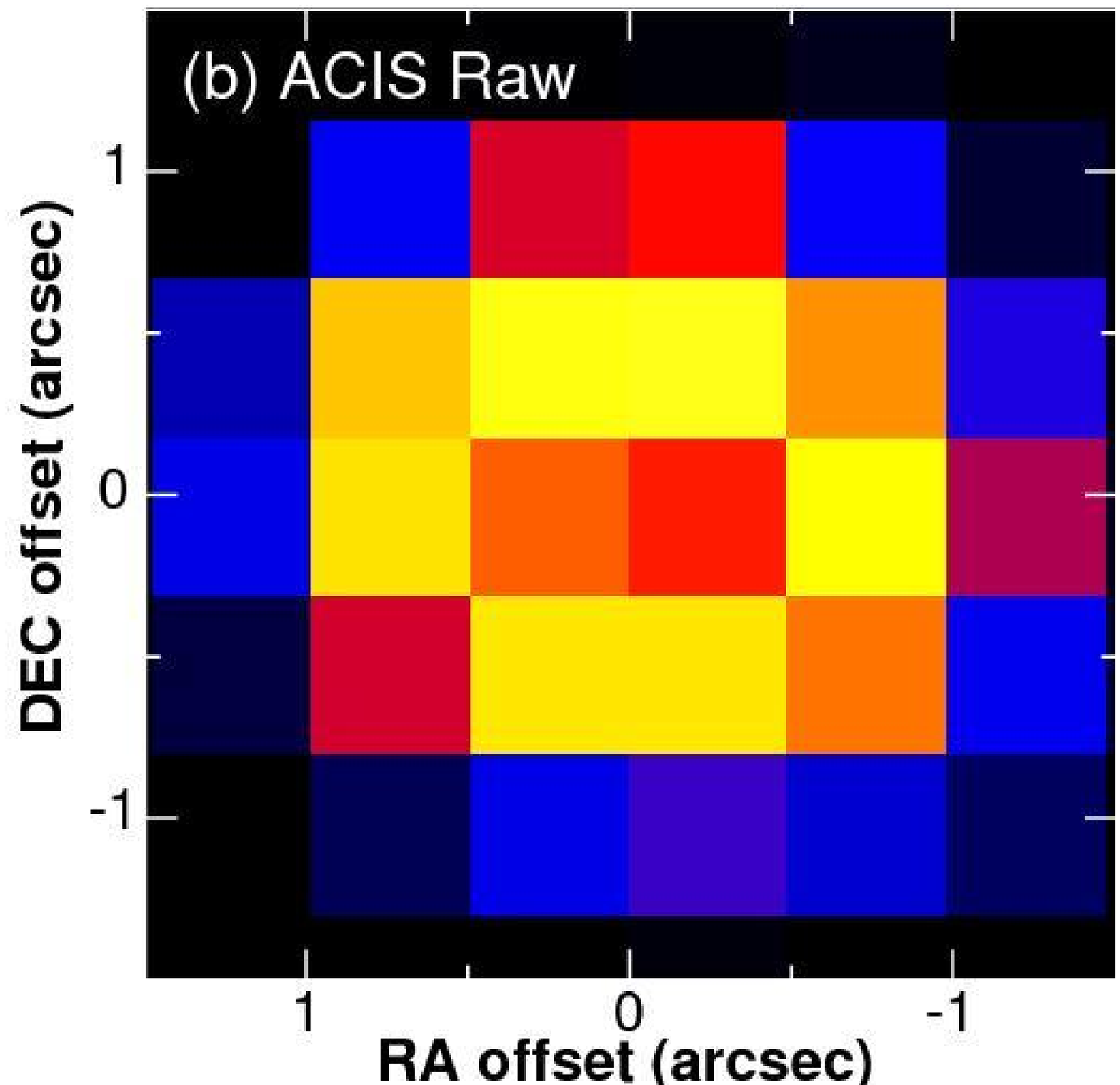}
 \end{minipage}
\hspace*{-3.8mm}
 \begin{minipage}[c]{0.3\textwidth}
  \includegraphics[height=53mm, bb=89 49 544 503, clip=]{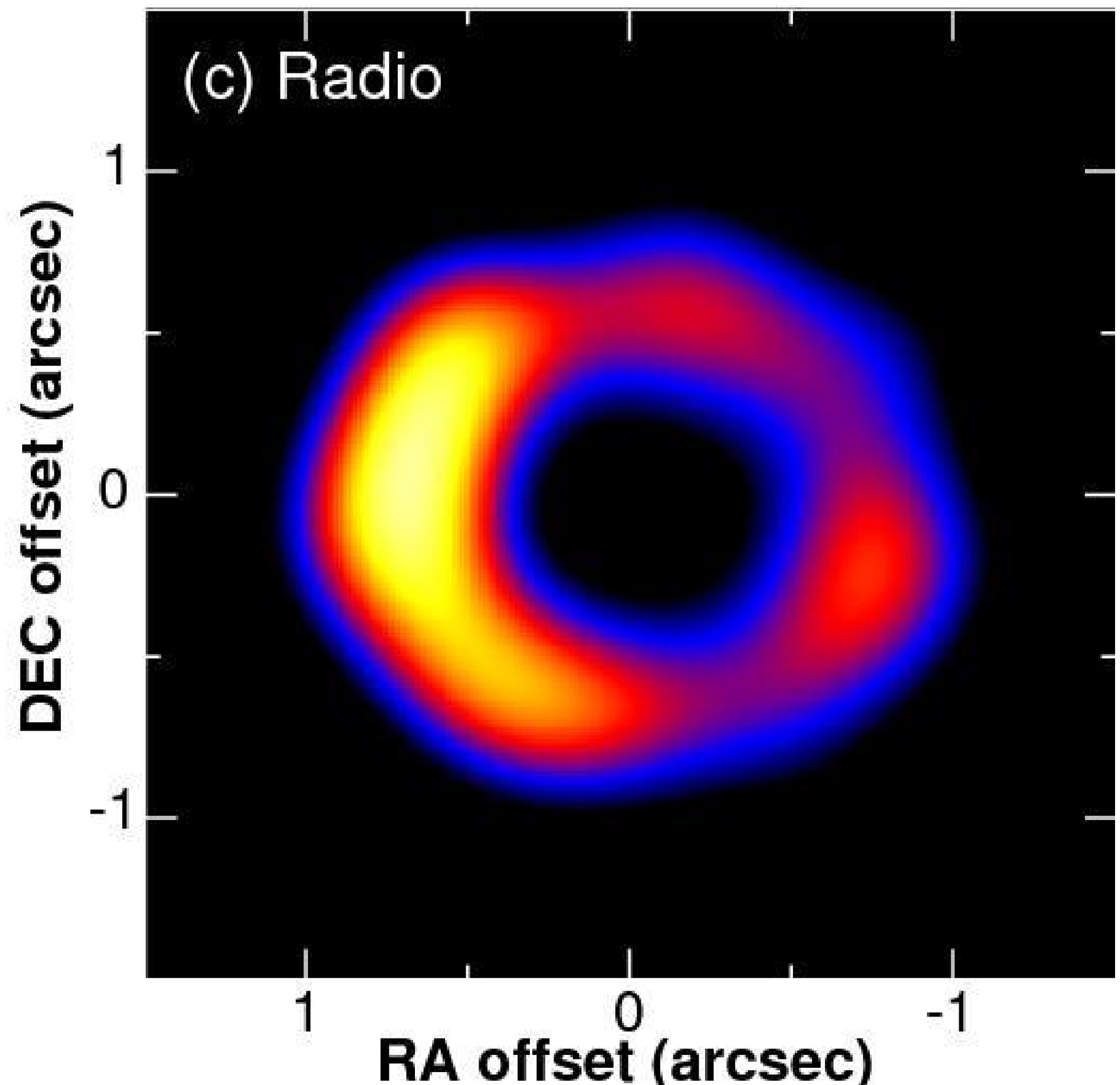}
 \end{minipage} \\[-1pt]
\hspace*{0.1mm}
 \begin{minipage}[c]{0.3\textwidth}
  \includegraphics[height=58.7mm, bb=5 1 544 504, clip=]{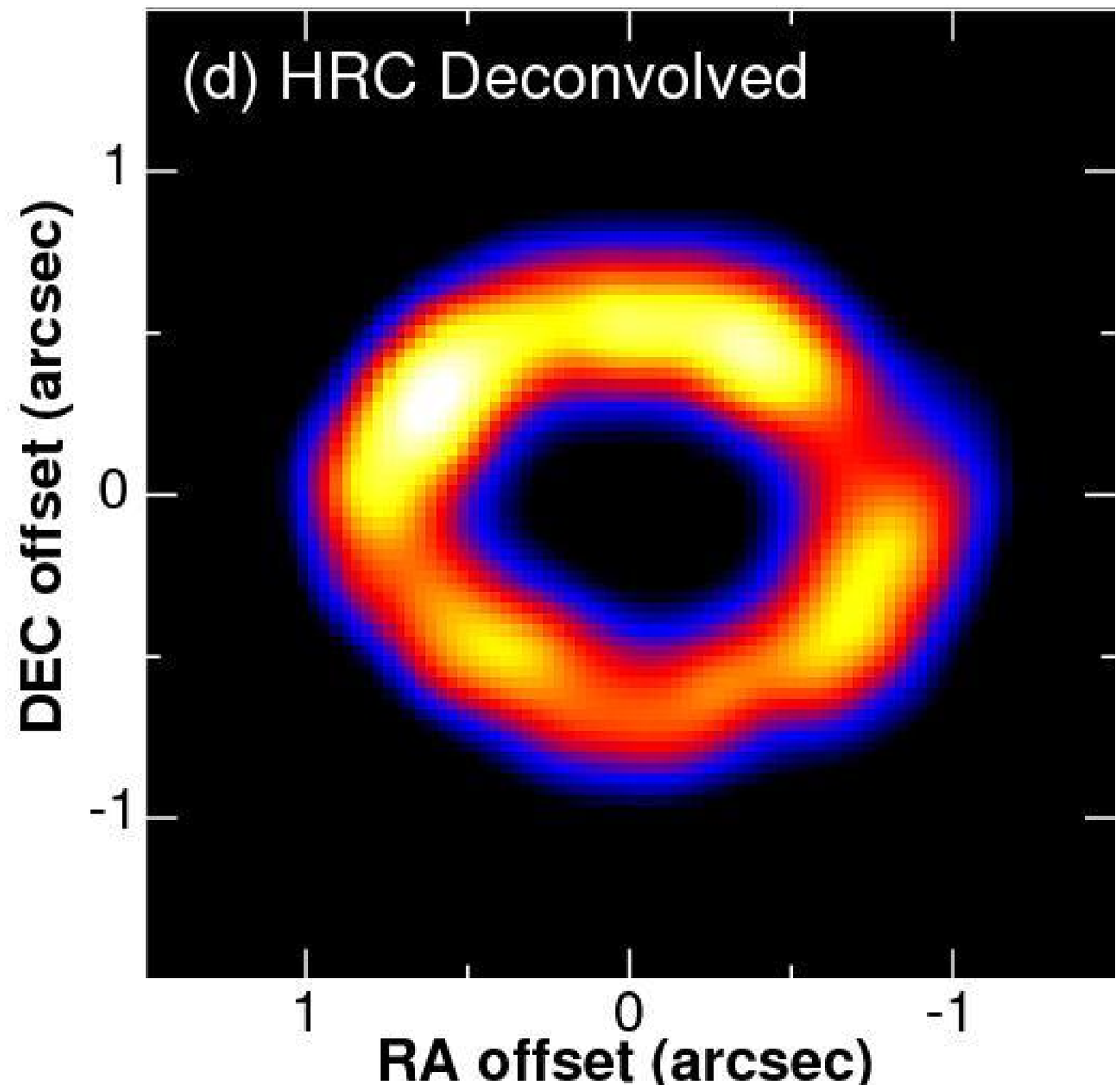}
 \end{minipage}
\hspace*{5.98mm}
 \begin{minipage}[c]{0.3\textwidth}
  \includegraphics[height=58.7mm, bb=89 1 544 504, clip=]{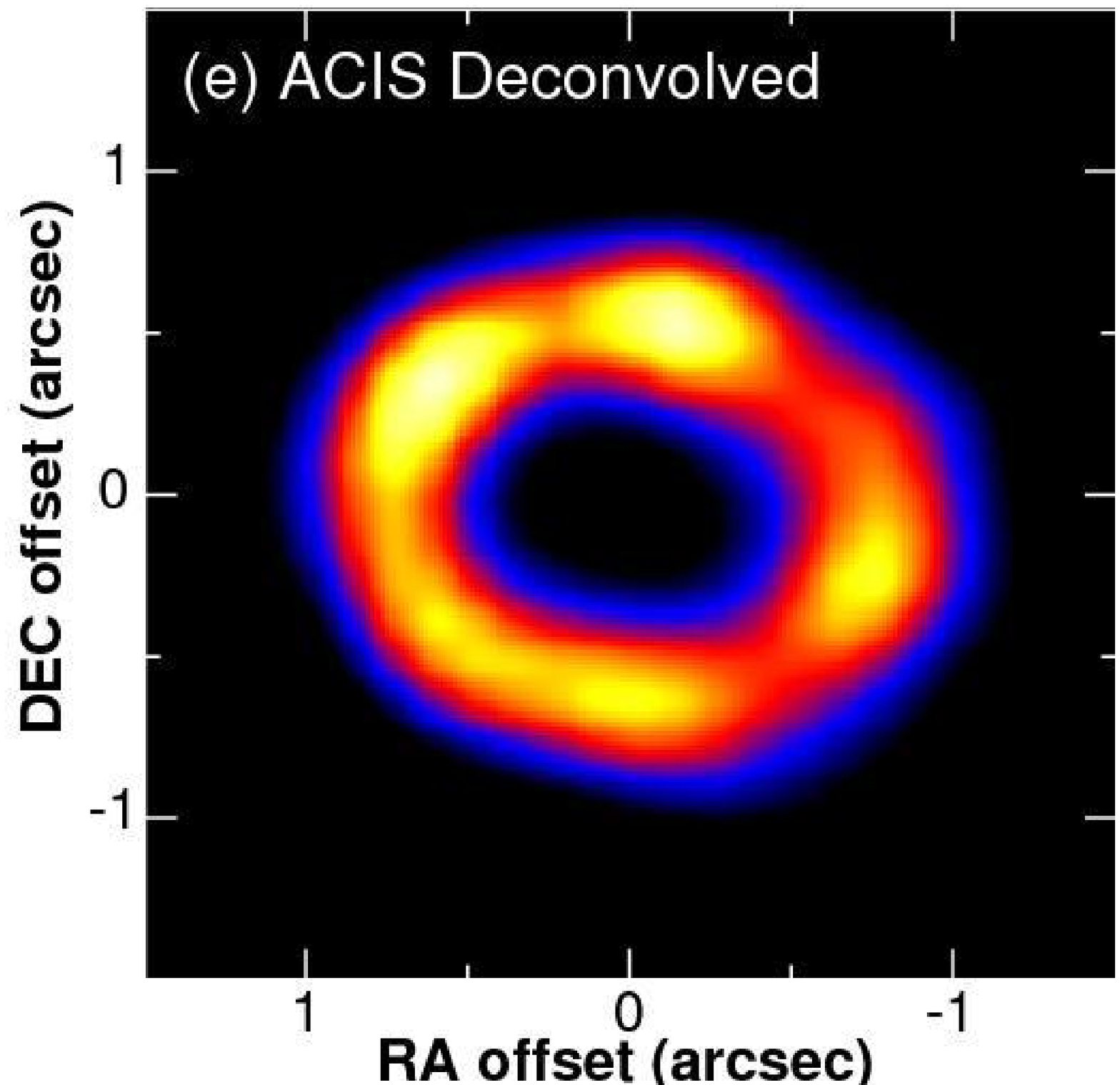}
 \end{minipage}
\hspace*{-3.8mm}
 \begin{minipage}[c]{0.3\textwidth}
  \includegraphics[height=58.7mm, bb=89 1 544 504, clip=]{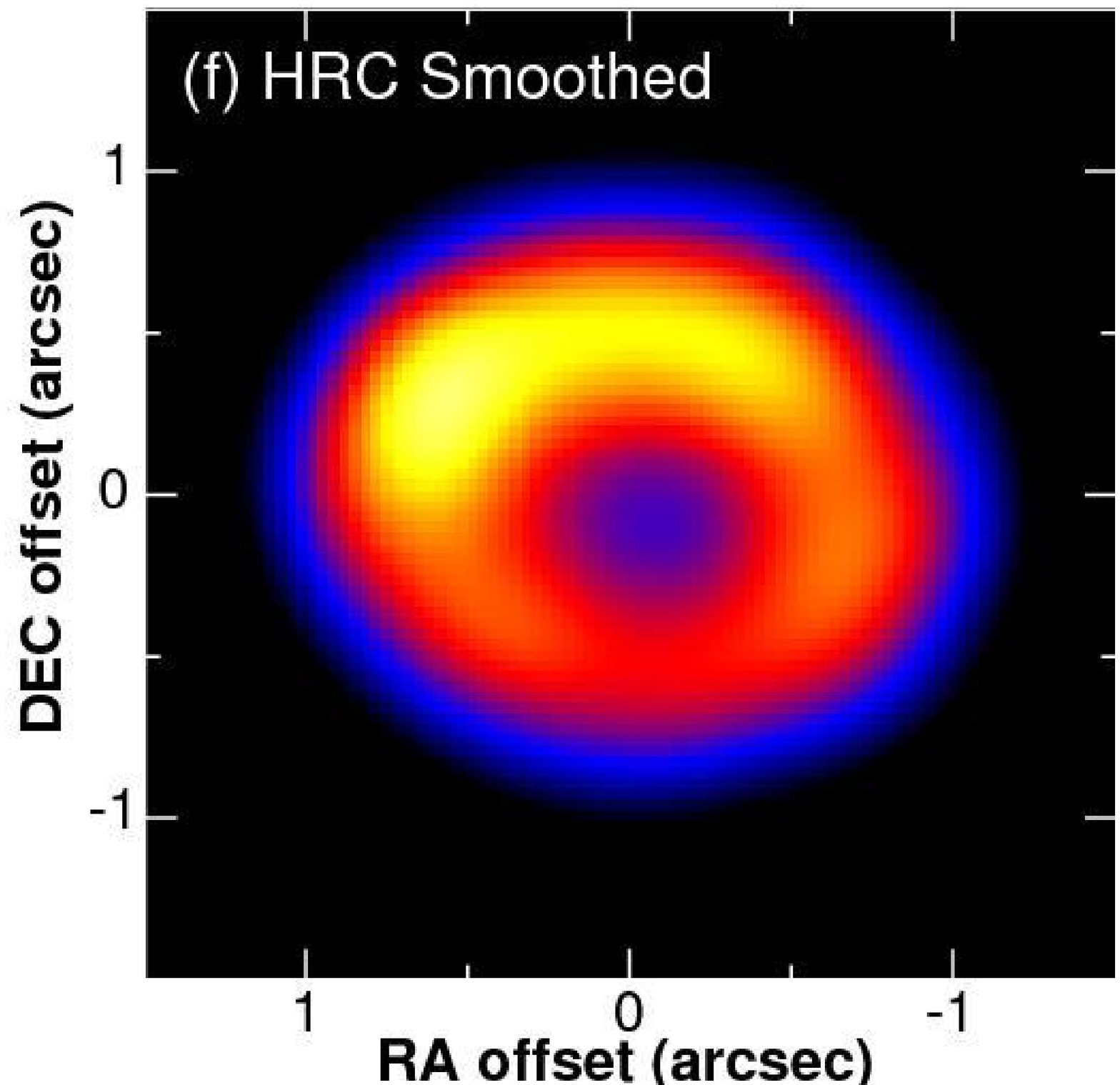}
 \end{minipage}
\caption{(a) Raw HRC data of SNR~1987A taken on 2008 Apr 28-29. The image
was binned into the HRC detector pixel and centered at
RA=$\mathrm{05^h35^m28^s}$, Dec=--69\arcdeg16\arcmin11.2\arcsec\ (J2000).
(b) Raw ACIS data taken on 2008 Jan 9. The image is in 0.3-8\,keV energy
range and was binned into the ACIS detector pixel. (c) Super-resolved radio
image at 9\,GHz taken by ATCA on 2008 Apr 23 \citep{ngs+08}. (d) Deconvolved
HRC image using the dataset shown in panel (a). (e) Deconvolved ACIS image
using the dataset shown in panel (b), in 0.3-8\,keV energy range \citep{rpz+09}.
(f) HRC data in panel (a) smoothed to 0.4\arcsec\ to match the
resolution of the radio image in panel (c). All panels are on the same spatial
scale.\label{f1}}
\end{figure*}

\section{Observation and Data Reduction}
Our \emph{Chandra} observations were carried out on 2008 Apr 28-29 (day
7736 since the SN explosion) using the HRC-I detector with a total exposure
of 46\,ks (ObsIDs \dataset[ADS/Sa.CXO#obs/09085]{9085} and
\dataset[ADS/Sa.CXO#obs/09851]{9851}). All the data reduction was
carried out using CIAO 4.1\footnote{\url{http://cxc.harvard.edu/ciao/}}
with CALDB 4.1.1. In addition to standard data processing, we applied
a screening algorithm \citep{mck+00} to reject the `bad events'
including non-X-ray background and mislocated events, thus improving the
data quality and boosting the signal-to-noise ratio. There were no
strong background flares during the exposure; hence all data were
included in the analysis.

\section{Spatial Analysis and Results}
The HRC image of SNR~1987A is shown in Figure~\ref{f1}a. This reveals
a ring-like morphology for the remnant with a subluminous region (a `hole')
at the center. The X-ray emission peaks in the northeast, and the north
rim is generally brighter than the south. A comparison to the raw ACIS
image at a similar epoch (2008 Jan 9) is shown in Figure~\ref{f1}b. The
HRC image resolves, for the first time, the remnant structure with direct
X-ray imaging, and reveals the central hole unambiguously. The entire
remnant has $(3.0\pm0.02)\times10^4$\,cts in the full energy range
(0.08-10\,keV) of the HRC and the background is negligible ($\sim0.1\%$),
implying a count rate of $0.66\pm0.01$\,cts\,s$^{-1}$.

To further improve the resolution of the HRC image, we applied a similar
deconvolution process as used by \citet{bmh+00}. The data were binned to a
quarter of the detector pixel in each dimension, then deconvolved using the
Lucy-Richardson algorithm \citep{luc74} with a
ChaRT/MARX\footnote{\url{http://cxc.harvard.edu/chart/}}-simulated
PSF at 1\,keV, where both the source spectrum and the detector response
peak.\footnote{\url{http://cxc.harvard.edu/proposer/POG/html/chap4.html}}
The resulting image after 30 iterations was smoothed to 0.2\arcsec\ 
resolution and is shown in Figure~\ref{f1}d. As a comparison, we also
showed in Figure~\ref{f1}e the deconvolution ACIS image from
\citet{rpz+09}. In both images, the remnant exhibits a similar annular
structure with sharp edges, and has semi-major and semi-minor axes
of $0.8\arcsec$ and $0.6\arcsec$, respectively. There are some very minor
difference in the extent of the bright regions between the two images,
but the brightness variation is mostly continuous and smooth around the
ring, and there is no obvious point source at the remnant center.

\begin{figure*}[ht]
\centering
 \begin{minipage}[c]{0.35\textwidth}
  \includegraphics[width=\textwidth]{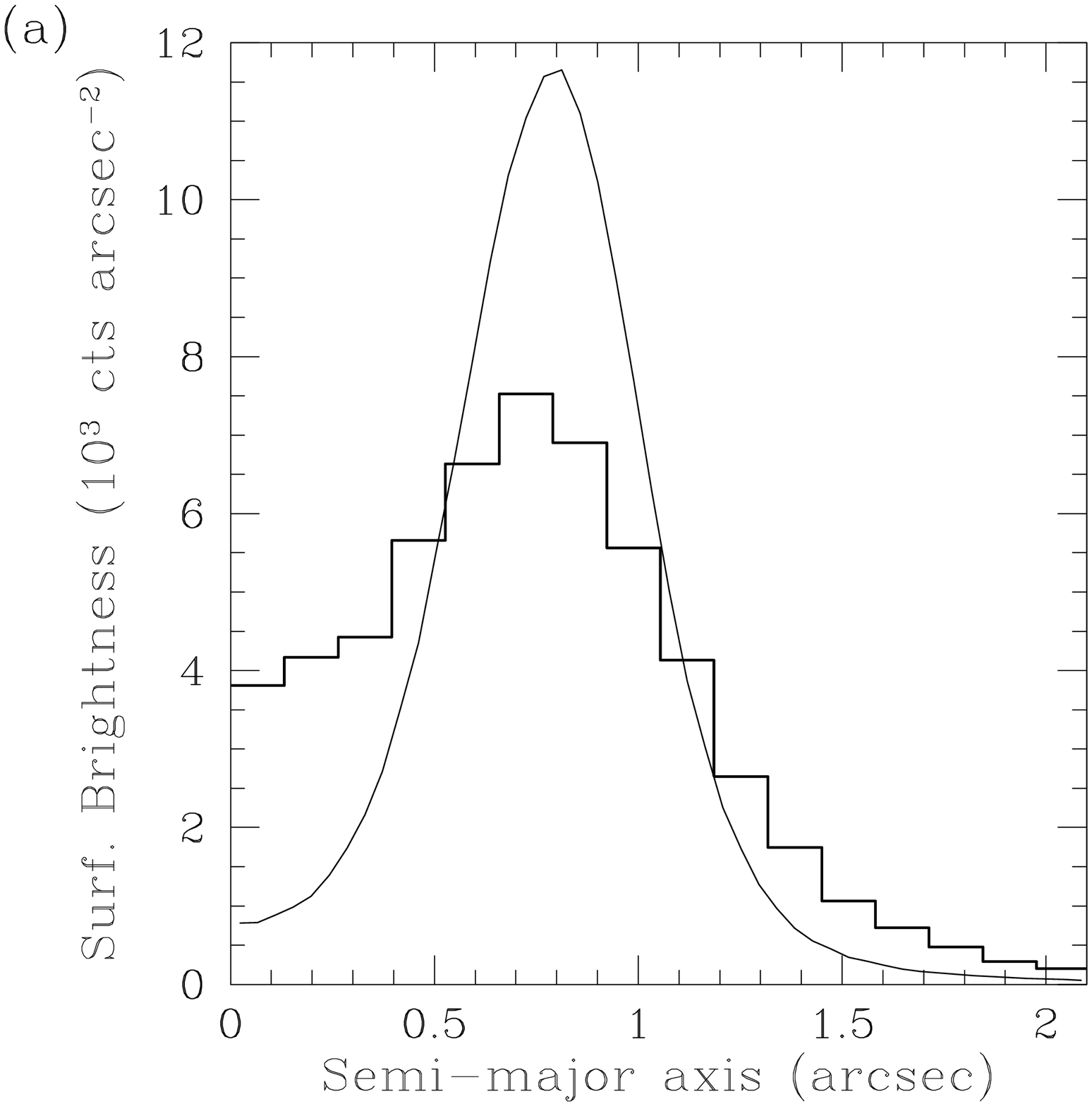}
 \end{minipage}
\hspace*{8mm}
 \begin{minipage}[c]{0.35\textwidth}
  \includegraphics[width=\textwidth]{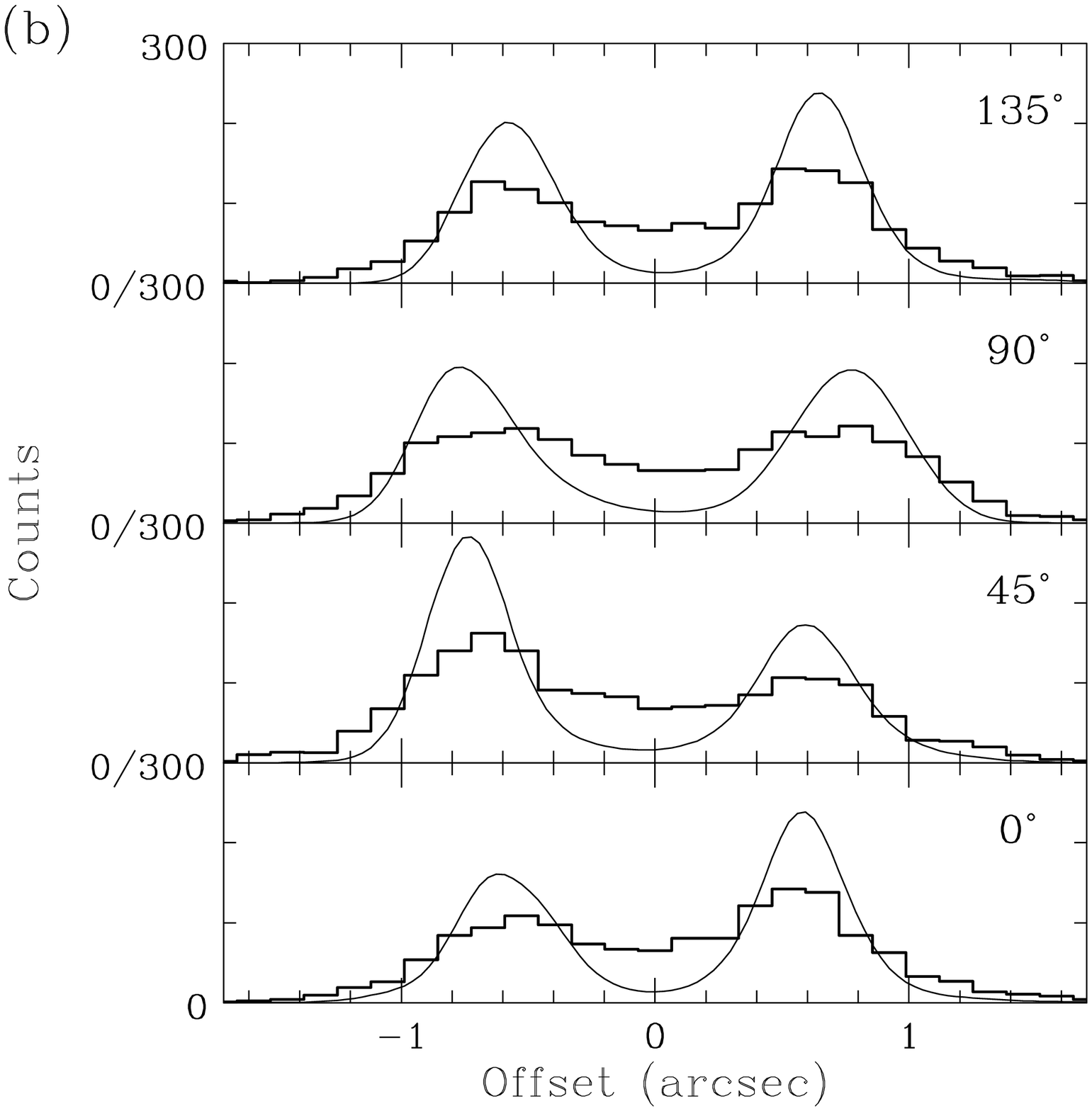}
 \end{minipage}
\caption{(a) Surface brightness profiles of SNR~1987A extracted from
elliptical annulus regions (see text). (b) Slices through the remnant
at different position angles (north through east). The offset is
radial from the center and is positive toward north and/or west. In
both plots, the raw and deconvolved HRC images are shown by the histograms
and lines, respectively. \label{f2}}
\end{figure*}

Figure~\ref{f1}c shows the super-resolved radio image of SNR~1987A at
9\,GHz, taken by the ATCA at a similar epoch \citep[2008 Apr 23;][]{ngs+08}.
To obtain a direct comparison between the radio and X-ray images, we
have regridded and smoothed the HRC data to the same resolution of
0.4\arcsec\ FWHM, and the resulting image is shown in Figure~\ref{f1}f.
The radio shell consists of two main lobes: a bright one in the east and
a fainter one in southwest; it has a similar size as the X-ray ring and
both emissions peak in the east. Also, there is some hint of an X-ray
counterpart for the southwestern radio lobe. However, except these
similarities, the radio and X-ray emissions exhibit substantially
different morphologies. While the former is mainly contributed by the
eastern lobe, the latter is brighest along the north rim, and neither
the overall bilateral pattern nor the strong east-west asymmetry in
the radio image are observed in the HRC data.

\begin{deluxetable*}{lcccccc}
\tablecaption{Best-fit Spatial Models to 2008 Apr 28-29 \emph{Chandra}
HRC Image of SNR~1987A\label{t1}}
\tablewidth{0pt}
\tabletypesize{\small}
\tablehead{\colhead{Model} & \colhead{Ring radius (\arcsec)} &
\colhead{Shell radius (\arcsec)} & \colhead{Thickness (\arcsec)} &
\colhead{$\chi^2$} & \colhead{dof \tablenotemark{a}} &
\colhead{$\chi^2_\nu$ \tablenotemark{b}}}
\startdata
Thin ring & $0.83\pm0.01$ & \nodata & $0.04\tablenotemark{c}$ & 1218& 4090 & 2.4 \\
Thick ring & $0.75\pm0.02$ & \nodata & $0.56\pm0.08$ & 942 & 4089 & 1.9 \\
Torus\tablenotemark{d} & \nodata & $0.82\pm0.02$ & $0.39\pm0.02$ & 884 & 4088 & 1.8 \\
Ring+shell\tablenotemark{e} & $0.82\pm0.03$ & $0.94\pm0.05$ & $0.05\tablenotemark{c}$ & 709 & 4088 & 1.4
\enddata
\tablenotetext{a}{The formal number of degrees of freedom (dof) depends on the image size
we are fitting, and they are likely overestimated (see text).}
\tablenotetext{b}{The reduced $\chi^2$ values, $\chi^2_\nu$, are estimated
with dof=500, as we argued in the text.}
\tablenotetext{c}{Fixed at 5\% of the corresponding radii during the fit.}
\tablenotetext{d}{The best-fit torus has a half-opening angle
$26\arcdeg\pm3\arcdeg$.}
\tablenotetext{e}{The fit suggest 18\% more counts in the
ring than the shell component.}
\tablecomments{The uncertainties quoted are statistical errors at 90\% confidence level.}
\end{deluxetable*}

We extracted surface brightness profiles for both raw and deconvolved HRC
images using elliptical annulus regions with aspect ratios fixed by the
system's orientation \citep[43.4\arcdeg\ to the line of sight;][]{pun07},
and the results are plotted in Figure~\ref{f2}a. The X-ray emission is
very faint at the remnant center and peaks at about 0.8\arcsec\ radius with
relatively sharp boundaries. The profile of the deconvolved image indicates
a width of FWHM 0.5\arcsec\ for the ring, but we note that the actual width
is likely smaller, since the deconvolved image has been smoothed.
Figure~\ref{f2}b shows slices through the remnant along different position
angles, illustrating a high degree of symmetry along the east-west direction.
We found that counts in the eastern and western halves differ only by
$5\%\pm3\%$. In contrast, the north-south asymmetry is more significant
($25\%\pm7\%$). It is partly due to the light travel time effect: the rapid
brightening of the remnant makes the north rim, which is currently 1.5\,light
year closer to the Earth than the south rim, appear brighter. Based on the ACIS
lightcurve, we estimated that this effect could result in at most 15\%
count difference between the northern and southern halves. However,
it is not enough to account for the observed value; therefore, we conclude
that the rest is intrinsic.  

Motivated by the images, we have carried out detailed spatial
modeling to capture the characteristic scales of the SNR. Simple
models were generated in 3D, corrected for light travel time
effects and projected onto the image plane according to the viewing
geometry. The resulting images were then convolved with the PSF
and fitted to the raw HRC data shown in Figure~\ref{f1}a using the
procedure described by \citet{nr08} with \citet{geh86} $\chi^2$
statistic. Since we are mostly interested in the global geometry,
we followed \citet{ngs+08} to account for the overall asymmetry
using a linear brightness gradient with arbitrary position angle
in the equatorial plane. It is worth noting that as compared to the
recent ACIS study \citep{rpz+09}, our fitting procedure involves no
resampling or deconvolution processes that may degrade the data
or introduce artifacts in the images. Therefore, it can provide
robust measurements.

\begin{figure*}[ht]
\centering
 \begin{minipage}[c]{0.3\textwidth}
  \includegraphics[height=53mm, bb=5 49 544 503, clip=]{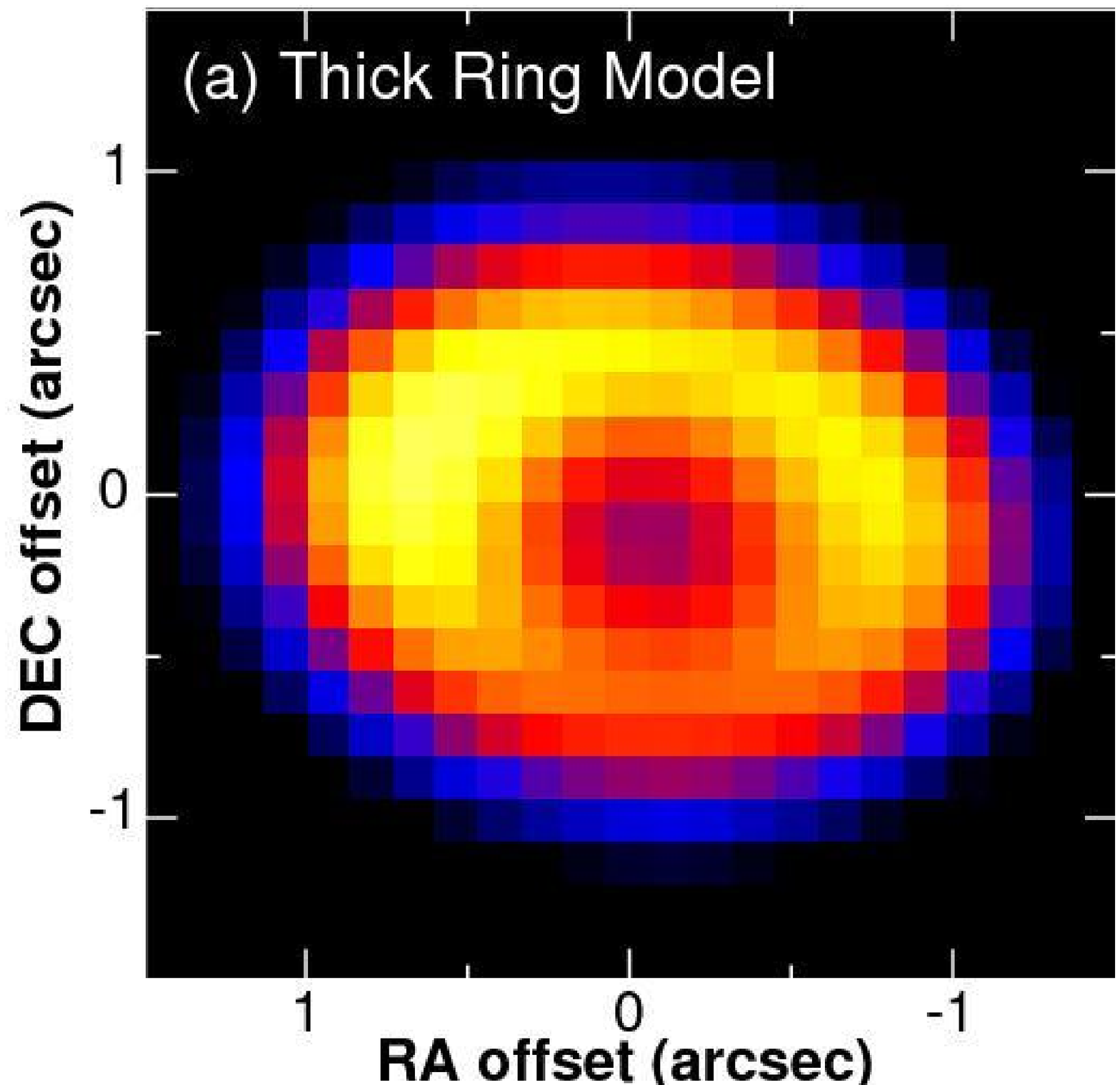}
 \end{minipage}
\hspace*{6.0mm}
 \begin{minipage}[c]{0.3\textwidth}
  \includegraphics[height=53mm, bb=89 49 544 503, clip=]{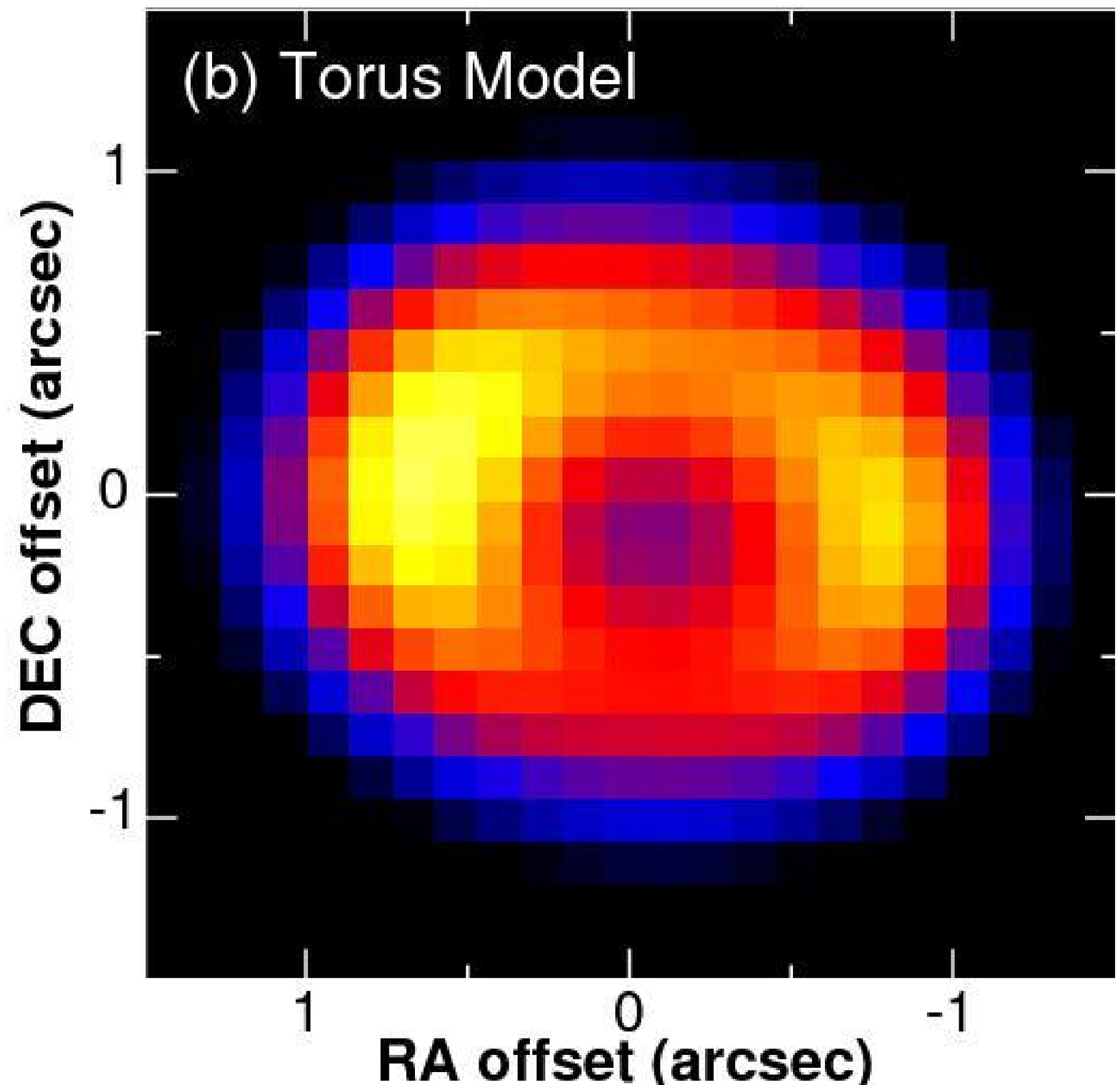}
 \end{minipage}
\hspace*{-3.8mm}
 \begin{minipage}[c]{0.3\textwidth}
  \includegraphics[height=53mm, bb=89 49 544 503, clip=]{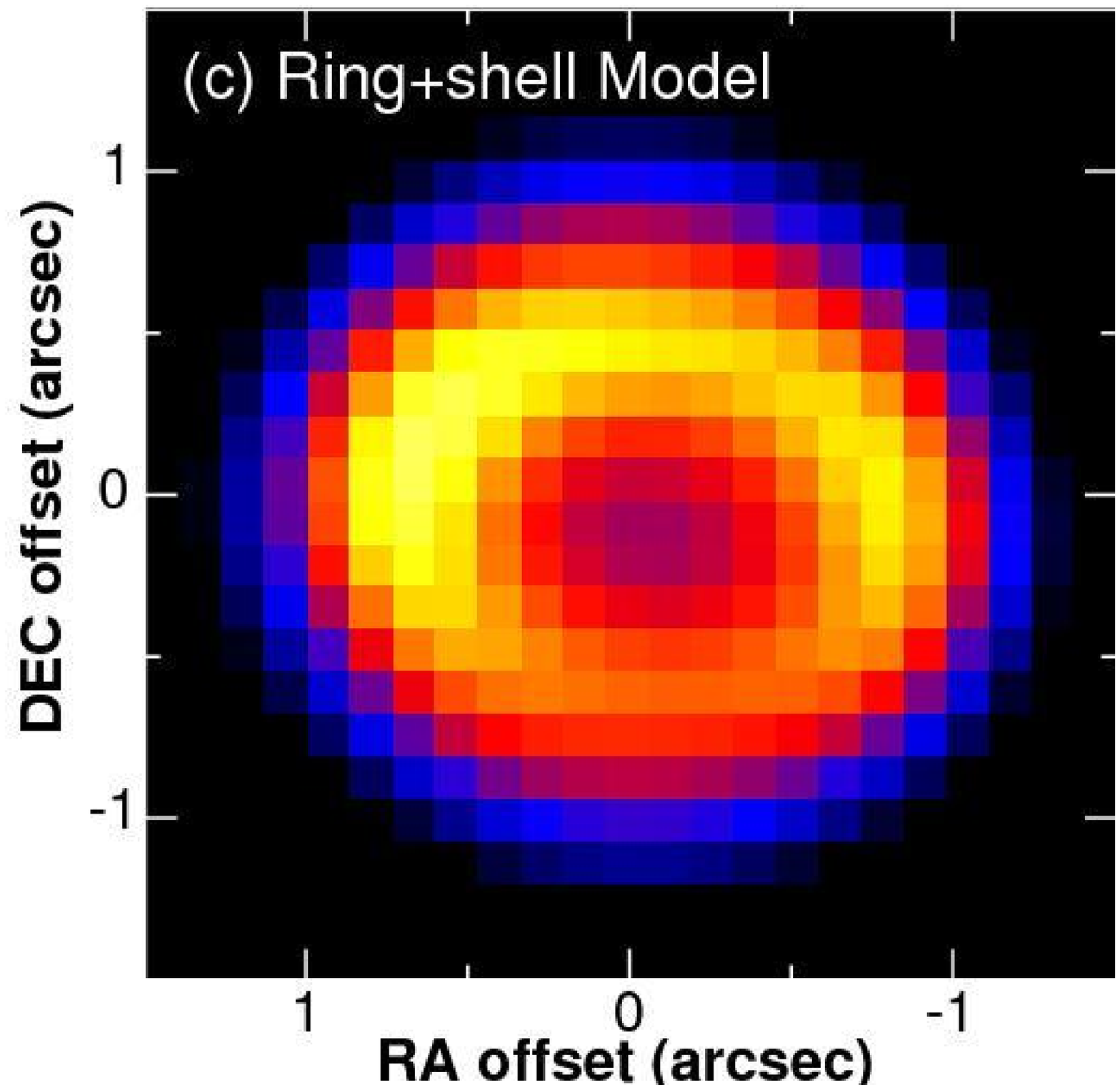}
 \end{minipage} \\[-1pt]
\hspace*{0.1mm}
 \begin{minipage}[c]{0.3\textwidth}
  \includegraphics[height=58.7mm, bb=5 1 544 504, clip=]{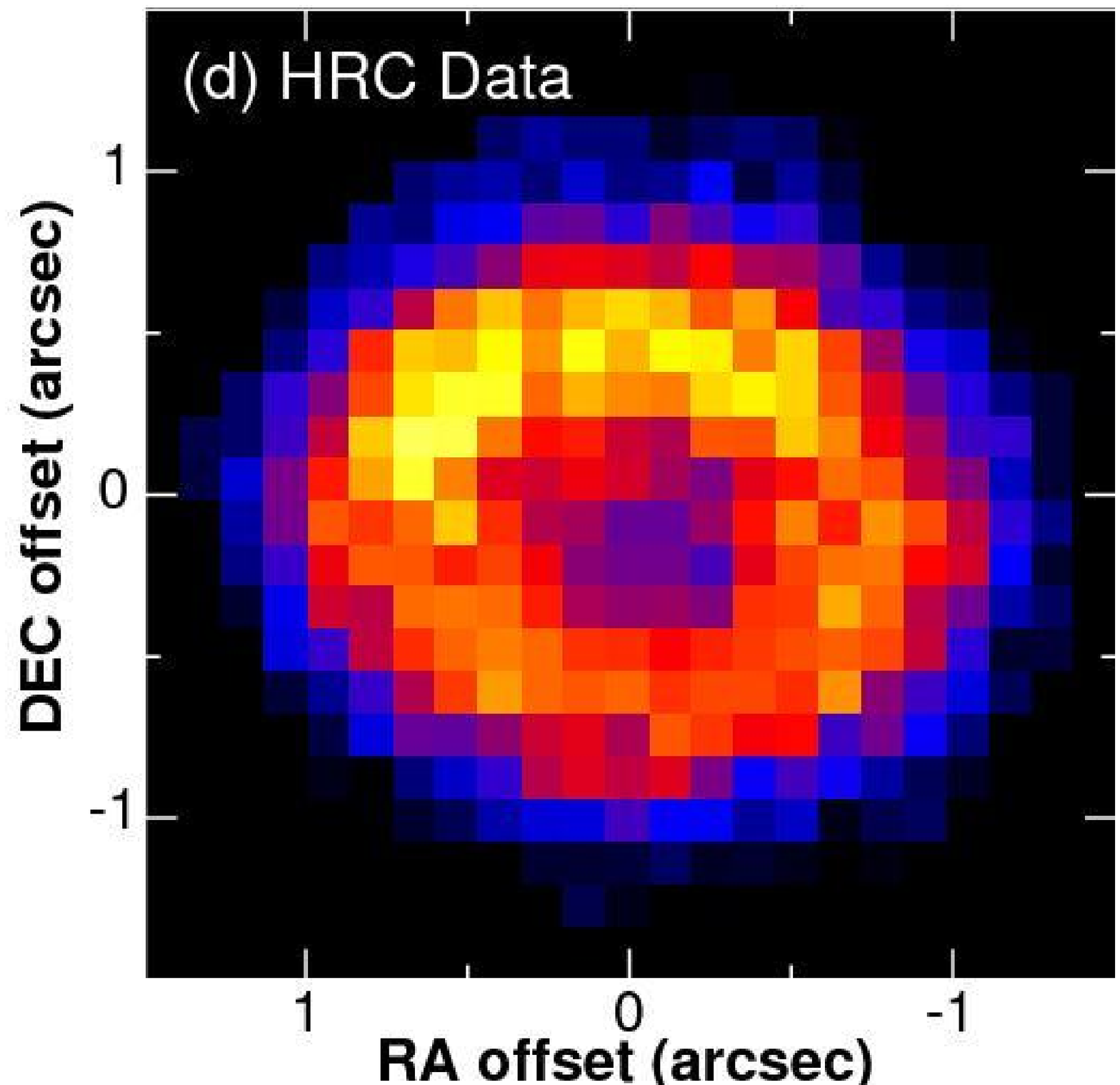}
 \end{minipage}
\hspace*{5.98mm}
 \begin{minipage}[c]{0.3\textwidth}
  \includegraphics[height=58.7mm, bb=89 1 544 504, clip=]{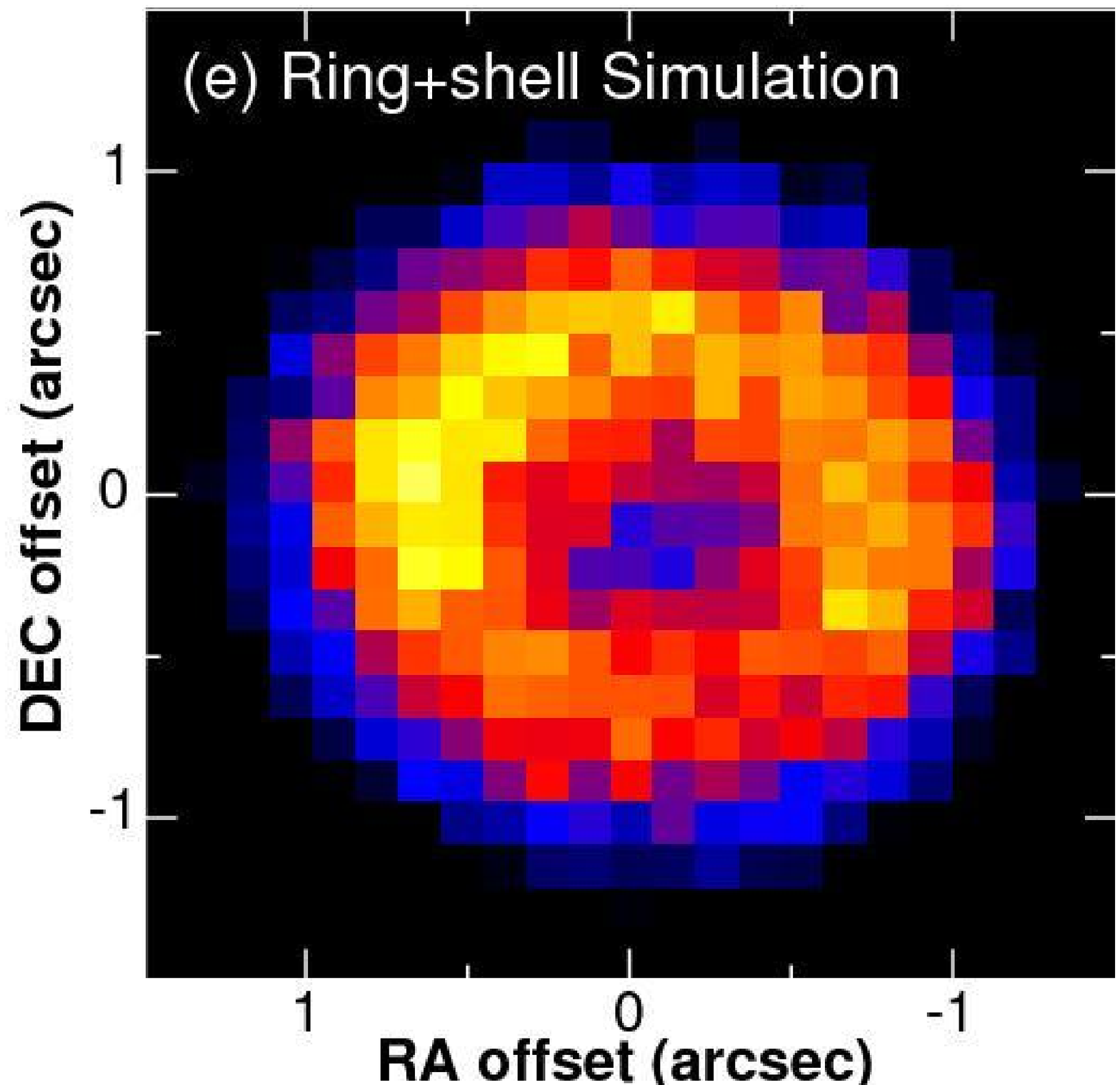}
 \end{minipage}
\hspace*{-3.8mm}
 \begin{minipage}[c]{0.3\textwidth}
  \includegraphics[height=58.7mm, bb=89 1 544 504, clip=]{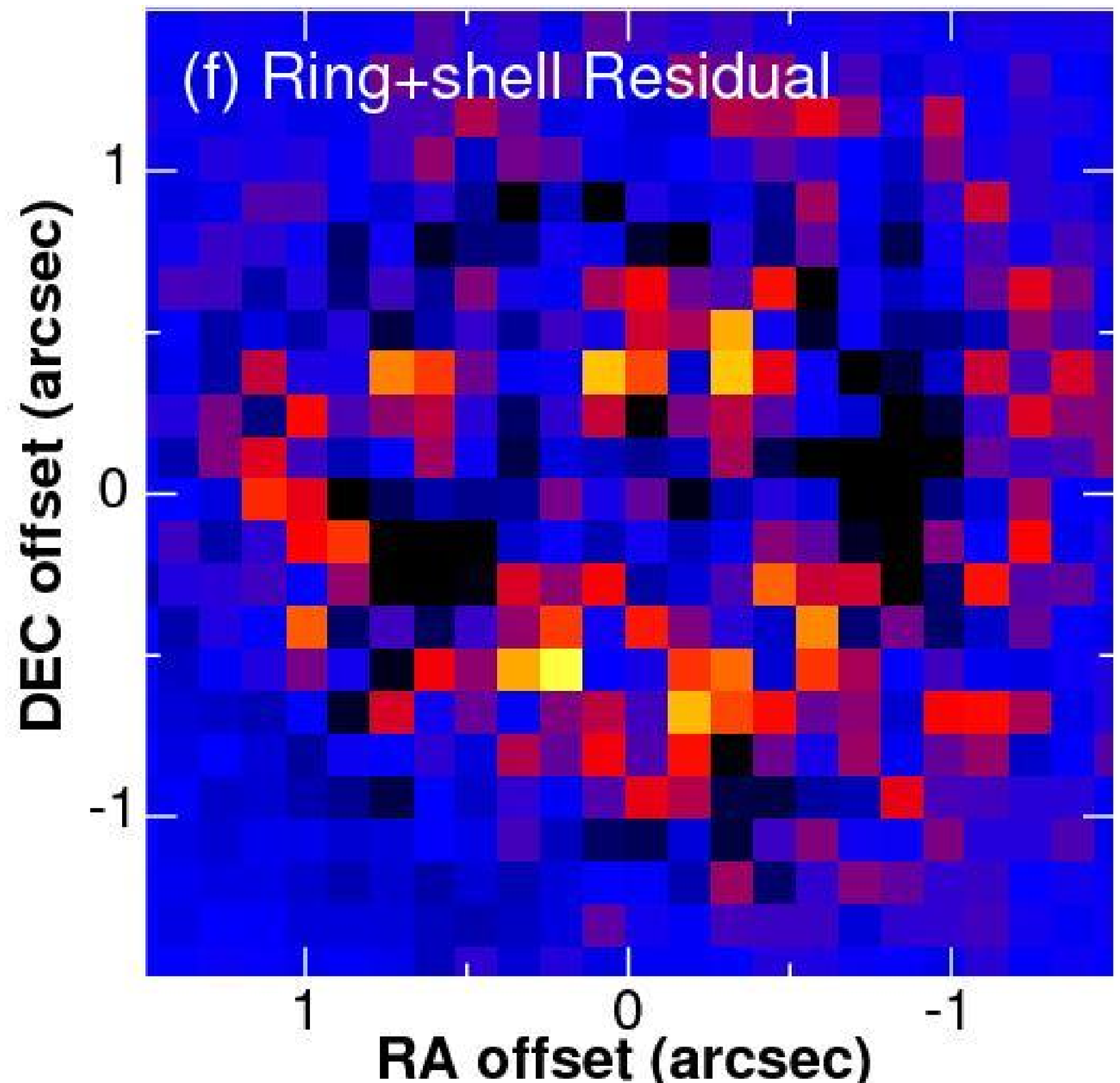}
 \end{minipage}
\caption{Best-fit spatial models for the HRC image of SNR~1987A: (a) a
thick ring, (b) a torus, (c) a thin ring plus a thin shell. All images are
convolved with the instrument PSF. (d) The HRC data, same as Fig.~\ref{f3}a.
(e) MARX simulation of the best-fit ring plus shell model from (c).
(f) Difference image between panels (d) and (c), showing the
residuals of the fit. All panels except (f) are on the same color range.
\label{f3}}
\end{figure*}

We have tried fitting various models, including a simple ring, an
equatorial belt torus \citep[see][]{ngs+08}, and a ring plus a shell.
The best-fit parameters are listed in Table~\ref{t1} with the corresponding
images shown in Figure~\ref{f3}. Note that the formal reduced $\chi^2$
values for the fits are likely underestimated, since the number of
independent pixels in the image we are fitting is much larger than the
source area. Better estimates of the reduced $\chi^2$ values could be made
using the source area of $\sim500$\,pixels. As shown in the table,
this gives more reasonable results. The uncertainties quoted in Table~\ref{t1}
are statistical errors with 90\% confidence level, estimated from Monte
Carlo simulations \citep{nr08}. We also followed their procedure to quantify
systematic errors due to features not captured by the simple models, such
as small-scale brightness fluctuations around the ring. This suggests a
systematic uncertainty of $\sim3\%$ for the radius. As indicated in
Table~\ref{t1}, different fits give similar radii of $\sim0.8\arcsec$,
consistent with the plot in Fig.~\ref{f2}a. We found that a thin ring fits
poorly to the data, but varying the ring's thickness significantly improves
the result. Our thick ring model is essentially same as the simple torus
used in the ACIS study of \citet{rpz+09}, and the radius and thickness we
obtained are in good agreement with their results. Our equatorial belt torus
model offers a better fit than a ring, and suggests a half-opening angle
$\sim26\arcdeg$, slightly smaller than that of the radio shell
\citep{ngs+08,psn+09}. However, the characteristic two-lobe structure of the
equatorial torus does not match the data (Fig.~\ref{f3}b). In terms of the
statistics, our best model is given by a thin shell plus a thin ring.
As shown in Figs.~\ref{f3}c \& \ref{f3}e, this model captures the overall
remnant morphology well and successfully reproduces the observed ellipticity.
The fit suggests 18\% more counts in the ring component than the shell.
While it is tempting to associate the latter with faster shocks at a higher
latitude, the physical scenario is likely to be more complicated and the
parameters in the fits could be highly degenerate. Further observations are
needed to confirm if a two-component spatial model is required.

Similar to \citet{ngs+08}, we added a point source at the center of the
best-fit model to estimate the flux limit from any possible central
object. We adjusted the point source flux and re-fit the model until the
$\chi^2$ value exceeded a certain level. This gives a detection limit
0.010\,cts\,s$^{-1}$ at 99\% confidence level over the energy range
0.08-10\,keV. For a $R^\infty=13$\,km radius\footnote{This is the redshifted
value as observed at infinity, same for the temperature and luminosity below.}
neutron star 51.4\,kpc away with a foreground neutral hydrogen column density
$N_{\rm H}=1.3\times10^{21}$\,cm$^{-2}$ \citep{zmd+09}, the flux limit
corresponds to a surface temperature $T^\infty=2.50$\,MK, or a bolometric
luminosity $L_{\rm bol}^\infty=4.7\times 10^{34}$\,ergs\,s$^{-1}$. More
physical estimates using neutron star atmosphere models \citep[e.g.][]{zps96}
suggest similar values. If the radiation from the central source is nonthermal
with a typical powerlaw index $\Gamma=1.5$, then the limit converts to a
luminosity of $7.0\times10^{34}$\,ergs\,s$^{-1}$ in 2-10\,keV energy range.
To have a fair comparison with previously published values, we followed
\citet{pzb+04} to ignore the statistical fluctuations of the underlying
remnant, and repeated the above exercise. This gives a 90\% confidence level
detection limit $3.3\times10^{34}$\,ergs\,s$^{-1}$ in the 2-10\,keV range, slightly
higher than the ACIS limit of $1.5\times10^{34}$\,ergs\,s$^{-1}$ \citep{pzb+04},
likely due to the brightening of the remnant, but more stringent than the
\emph{XMM-Newton} limit $5\times10^{34}$\,ergs\,s$^{-1}$ obtained from
spectral analysis \citep{slg+05}.

\section{Discussion}
\subsection{X-ray Morphology of SNR 1987A}
The HRC observations of SNR~1987A have provided a direct X-ray image that
confirms the remnant structure as suggested by previous ACIS observations
using the deconvolution technique \citep[][and references therein]{rpz+09}.
As the image shows, the entire X-ray ring has been lit up, and its
size and ellipticity are in good accord with the optical inner ring.
This gives direct evidence that the blast wave has already encountered
the dense CSM all around the ring and is now interacting with the main
body of the ring. The deconvolved HRC and ACIS images are nearly identical,
except for some very minor differences in the brightness distribution, which
could be attributed to the different energy response of the instruments. 
\citet{zmd+09} reported a strong energy dependence of the east-west
asymmetry in the ACIS data. The eastern half is dominated by hotter
plasmas behind fast shocks, therefore it is brighter at higher energies.
Conversely, the western half has cooler plasmas, slower shocks and is
brighter at lower energies. These two contributions could possibly
cancel out in the HRC image, resulting in the high degree of symmetry
observed along the east-west direction. We are unable to verify this because
the HRC data do not possess any spectral information. As the shocks further
propagate and decelerate in the dense CSM, we expect the degree of symmetry
to vary in future observations.

The poor spatial correlation between the soft X-ray and radio morphologies
suggests that they may originate from different physical processes. While
the X-rays trace the SN blast wave, the radio emission is believed to originate
between the forward and reverse shocks \citep{mgs+05,ngs+08}. Given the
young age of SNR~1987A, both shocks should have similar radii
\citep[e.g.][]{tm99}, hence, one would expect the X-ray and radio
shells to have a comparable size. A direct comparison between the HRC
and radio images in Figure~\ref{f1} seems to support this picture. Nonetheless,
previously reported radii for the X-ray shell have always been 10-15\% smaller
than for the radio shell \citep{ngs+08,rpz+09}. The discrepancy could be due
to different measuring techniques \citep{gsm+07,rpz+09}, or due to
systematic errors in the measurements. A better way to compare the
sizes is from the outer edge of the shells. This can avoid complication
by the inner ring or projection effects. After accounting for the shell's
thickness, our best-fit model of the HRC image has an outer radius of
$0.96\arcsec$. Similar values are obtained from other models (see
Table~\ref{t1}), and it also seems consistent with the ACIS study
\citep{rpz+09}. All these indicate that our estimate is robust, thus, we
conclude that the X-ray shell has an outer radius of $0.96\arcsec\pm0.05
\arcsec\pm0.03\arcsec$, with the uncertainties corresponding to
statistical (90\% confidence) and systematic errors, respectively. This
is very similar to the outer radius $0.92\arcsec\pm0.06\arcsec$ of the
radio shell \citep{psn+09}, thus confirming the physical picture discussed
above. Although these values appear to be slightly larger than the optical
inner ring of radius 0.81\arcsec-0.86\arcsec \citep{bkh+95,plc+95}, they
unlikely represent the radius of the transmitted shock inside the dense CSM,
since which should have a very slow propagation speed according to some
hydrodynamic simulations \citep[e.g.][]{dwa07}. Therefore, we believe that
our results could correspond to the projected size of fast shocks traveling
in a lower density environment at higher latitudes.

\subsection{X-ray Limit on a Central Source}
In the absence of fast cooling mechanisms, standard theories suggest a
bolometric luminosity $L_{\rm bol}^\infty \ge 5.0\times10^{34}$\,ergs\,s$^{-1}$
(corresponding to a surface temperature
$T^\infty\approx2.54$\,MK) for a 21-year-old neutron star \citep{yp04,sy08}.
Our flux limit rejects any unobscured neutron stars hotter than this
at 99\% confidence. \citet{gcc+05} argued that a neutron star
obscured by dust is unlikely, since the IR flux is consistent with the
radioactive decay of $^{44}$Ti in the ejecta and no other energy source is
required. The same conclusion also holds for the recent IR observations
\citep{bdd+06}. On the other hand, for a neutron star with short thermal
relaxation time due to superfluidity, quark matter, or anomalously high
thermal conductivity of the crust, the thermal emission from the surface
could have already been too low to detect at an age of 21 \citep{sy08,cch+09}.

X-ray emission from a young pulsar is often dominated by nonthermal
radiation from the magnetosphere and from the surrounding pulsar wind
nebula (PWN). The nonthermal flux limit we derived rules out energetic
pulsars such as the Crab or PSR~B0540--69, but a fainter pulsar/PWN system
similar to 3C~58 would still escape detection \citep[see][]{kp08}. These
authors also gave a general correlation between nonthermal luminosity and
spin-down power $\dot{E}$ of young pulsars, which places a conservative
limit $\dot{E}<3\times10^{37}$\,ergs\,s$^{-1}$ for the unseen pulsar in
SNR~1987A. Based on the non-detection, \citet{oa04} argued that the pulsar,
if it exists, should have either a very strong or very weak magnetic field.
While the former case would put the star into the magnetar regime, our
detection limit is incompatible with the X-ray luminosities of nearly all
known magnetars, even in quiescence \citep{wt06,mer08}. For the weakly
magnetized neutron star case, one candidate is a central compact object
(CCO), which is a soft X-ray source found inside a SNR, and could possibly
be a neutron star born with a weak magnetic field \citep[see review by][]{gh08}.
It has been proposed that SNR~1987A may harbor a CCO \citep[e.g.][]{man07,
hgc+07}; however, the current observations do not provide any useful
constraints on this scenario, unless CCOs have similar cooling properties
as discussed above. Finally, we note that \citet{gcc+05} considered different
accretion models and used the UV flux limit to rule out spherical accretion
and a slim disk, where the disk thickness is non-negligible. They also placed
stringent constraints on thin-disk models. Although it is beyond the scope of
this paper to reproduce their modeling, given the similar order of our X-ray
limit, we believe that our results unlikely improve their constraints
substantially.

\section{Conclusion}
We have presented a detailed analysis of \emph{Chandra} HRC observations of supernova
remnant 1987A. The remnant has a similar size as the optical inner ring,
and its morphology is statistically best-fitted by a thin ring plus a thin
shell. This provides direct evidence that the dense inner ring of circumstellar
medium has been overtaken the supernova blast wave. Our spatial modeling also
indicates a similar size between the X-ray-- and radio-emitting regions,
confirming the picture that the supernova forward and reverse shocks are
closely located. The X-ray flux limit on any possible central source rejects
a young neutron star, unless it has some fast cooling mechanisms or is
obscured by a small accretion disk.

\acknowledgements
We thank Svetozar Zhekov and Dima Yakovlev for useful discussions.
C.-Y.N. and B.M.G. acknowledge the support of the
Australian Research Council through grant FF0561298.
This work was supported in part through NASA Contract NAS 5--38248.
{\it Facilities:} \facility{CXO (HRC)}


\end{document}